\documentclass[preprintnumbers,amsmath,amssymb,aps]{revtex4}

\usepackage{graphicx}
\usepackage{multirow}
\usepackage{tabularx}

\preprint{DESY 07-125}
\preprint{COLO-HEP-529}

\newcommand{\bee}{\begin{equation}}
\newcommand{\ee}{\end{equation}}
\newcommand{\beea}{\begin{eqnarray}}
\newcommand{\eea}{\end{eqnarray}}

\def\Tr{{\rm Tr}}

\def\hx{\hat{x}}
\def\hy{\hat{y}}
\def\hm{\hat{m}}
\def\hmu{\hat{\mu}}
\def\hd{\hat{\delta}}

\begin{document}

\title{Parameters of the lowest order chiral Lagrangian from fermion eigenvalues}
\author{Thomas DeGrand}
\affiliation{
Department of Physics, University of Colorado,
Boulder, CO 80309 USA
}
\author{Stefan Schaefer}
\affiliation{
NIC, DESY,
Platanenallee 6,
D-15738 Zeuthen,
Germany
}

\begin{abstract} Recent advances in  Random Matrix Theory enable one to
determine the pseudoscalar decay constant from the response of eigenmodes
of quenched fermions to an imaginary isospin chemical potential.
We perform a pilot test of this idea,
from simulations with two flavors of dynamical overlap fermions.
\end{abstract}
\maketitle

\section{Introduction}
The leading-order chiral effective Lagrangian
\bee
{\cal L}_{\rm eff} = \frac{F^2}{4} \Tr (\partial_\mu U \partial^\mu U^\dagger)
- \frac{\Sigma}{2} \Tr[ {\cal M}(U+U^\dagger)]
\label{calL}
\ee
is characterized by two coupling constants, $F$ and  $\Sigma$. While they
are fundamental from the point of the chiral Lagrangian, their
values may be computed from the parent theory, QCD.
This is a long-standing goal of lattice simulations.

$F$ and $\Sigma$ are not measured directly in experiment: the values
of the decay constant and condensate for nonzero quark mass are computed from the
 chiral Lagrangian through
familiar functions of $F$ and $\Sigma$, plus contributions from higher order
terms in the chiral Lagrangian.
Lattice simulations regularly present results for
 the mass-dependent quantities. Since it is not so easy to do simulations
 directly at zero quark mass (or at the physical up or down quark masses),
the analysis of simulations  typically involves fits of lattice data at several
values of the quark mass  to
chiral perturbation theory formulas. $F$ and $\Sigma$ are
parameters determined as part of the fit.

Many techniques for extracting the mass dependent parameters, 
or $F$ and $\Sigma$ themselves,  have been devised.
Most of them require studying the asymptotic behavior of
correlation functions, and hence, need larger simulation volumes.
However,
in the epsilon regime ($m_\pi L  \ll 1$ but $\Lambda L \gg 1$ for any other
nonperturbative energy scale $\Lambda$), 
 properties of the chiral Lagrangian are encoded
in the spectrum of low-lying eigenvalues of the QCD Dirac operator in
 a finite
volume, whose distribution can be predicted by Random Matrix
Theory (RMT)~\cite{Shuryak:1992pi,Verbaarschot:1993pm,Verbaarschot:1994qf}.
Many groups\cite{
Berbenni-Bitsch:1997tx,Damgaard:1998ie,Gockeler:1998jj,
Edwards:1999ra,Giusti:2003gf,Follana:2005km,Berbenni-Bitsch:1998sy,
Damgaard:2000qt,
DeGrand:2006uy,
DeGrand:2006nv,
Lang:2006ab,
Fukaya:2007fb,
Fukaya:2007yv}
 have computed $\Sigma$ from fits to the
distributions of low-lying eigenmodes of the Dirac operator.
This seems to require less computer power for equivalent accuracy\cite{McNeile:2005pd}
 than ``asymptotic'' methods, and also seems to involve  fits
to data with fewer free parameters.

 A method to compute $F$ using eigenmodes was recently devised by
  G.~Akemann, P.~H.~Damgaard, J.~C.~Osborn and K.~Splittorff \cite{Akemann:2006ru}.
They imagine doing simulations in the epsilon-regime with two dynamical flavors,
and then considering the spectrum of a partially quenched theory in which two non-dynamical
flavors are coupled to an imaginary isospin chemical potential. The correlator
of the ordinary eigenvalues with those of the quenched ones involves the product of
the isospin chemical potential and $F$.
 The required eigenmode
distributions are  easily computed from an ordinary dynamical ensemble.
This paper is a pilot study of the technique at  lattice spacings which are small enough
that one can imagine quoting a continuum number. 

We can almost complete the calculation. The missing ingredient is actually analytic:
We are doing the calculation in finite simulation volume. We know that finite volume
alters the effective value of the condensate (said differently, the condensate appears
in all relevant formulas multiplied by a volume-dependent coefficient) \cite{Gasser:1987ah}.
 We expect that
$F$ is subject to similar modification. The precise calculation of this connection
does not appear to have been performed, and appears to require more knowledge of
chiral perturbation theory than we have. We return to this point below.

There have been two generations  of previous variations of this idea, Refs.
\cite{Damgaard:2005ys} and \cite{Damgaard:2006pu,Damgaard:2006rh}. These authors
 did simulations in which the isospin chemical
potential was included in the action. This is of course much more expensive than
what we are doing here.

We performed the calculation with 
lattice fermions which support the full $SU(N_f)\otimes SU(N_f)$ chiral symmetry
(including the anomalous singlet current and the index theorem), 
overlap fermions  \cite{Neuberger:1997fp,Neuberger:1998my} .
We  find doing overlap simulations to be reasonably straightforward.
RMT makes predictions for QCD in fixed topological sectors and its
predictions are for the pure imaginary eigenvalues of an anti-Hermitian Dirac operator.
Both of these requirements are met unambiguously (modulo cutoff effects) by overlap fermions:
the index theorem as realized by the Ginsparg-Wilson relation \cite{Ginsparg:1981bj} gives 
the topology and the usual definition of the condensate
as $\Sigma = (1/Z)\partial Z/\partial m$ stereographically maps overlap eigenmodes onto
the imaginary axis \cite{Niedermayer:1998bi}.

Typical high-statistics supercomputer collaborations (for example Ref.
\cite{Bernard:2006wx}) quote $f_\pi$ with 0.3 per cent errors (but slightly greater errors
 for $F$;
see below).
Ref. \cite{Damgaard:2006pu}, which extracted $F$ from eigenmodes 
(the isospin chemical potential is included in the simulation)
 has an 0.7 per cent error with about $10^3$ samples. This was done deep in strong coupling
so they did not attempt to make a continuum prediction.
We will not come close to this amount of accuracy due to lack of statistics, but our results
may not be uninteresting. Finding 
$F$ does not seem necessarily to be a supercomputer project.

If we were doing simulations in the real world of three light flavors,
what would we expect?
Two representative continuum chiral perturbation theory fits to data give
$F=88$ MeV   \cite{Gasser:1983yg}
or
$F=86.2\pm 0.5$ MeV  \cite{Colangelo:2003hf}.
Our simulations are for two dynamical flavors and so strictly speaking
we should not be talking about experimental comparisons. However, we will be unable to
resist doing this.

The rest of the paper is organized as follows: We first review the observables we are going
to compute and the corresponding Random Matrix predictions.
Then we discuss the systematic and statistical uncertainties of this simulation, focusing
on the corrections due to the finite volume. After that, the setup and results of 
our actual simulation will be given.

\section{Theoretical Background}
The key to the calculation is the connected
correlation function between eigenvalues of the Dirac
operator computed in the presence of a quenched chemical potential $\lambda_j$, and ordinary
eigenvalues computed at zero chemical potential, $\lambda_i$
\bee
\rho_{(1,1)}^{(2)\,conn}(x,y) =
\big \langle \sum_i \delta(x-\lambda_i) \sum_j \delta(y-\lambda_j)\big \rangle
- \big\langle \sum_i \delta(x-\lambda_i)\big \rangle\big \langle \sum_j \delta(y-\lambda_j)\big \rangle  .
\ee
Eq.~(3.49) of Ref.~\cite{Akemann:2006ru} gives the RMT prediction
for this correlator
in terms of the usual rescaled variables $\hx=\lambda_i \Sigma V,\hy=\lambda_j \Sigma V$,
 the rescaled mass
$\hm = m_q\Sigma V$,
and the rescaled isospin chemical potential $\hmu = \hd  = \mu F \sqrt{V}$.
$V$ is the volume.
\bee
\rho_{(1,1)}^{(2)\,conn}(\hx,\hy) =
\hx\hy \ \frac{
\left|\begin{array}{ccc}
 \mathcal{I}^+(\hy,i\hm) & J_\nu(i\hm) & i\hm J_{\nu+1}(i\hm) \\
 \mathcal{I}^+(\hy,i\hm)' & J_\nu(i\hm)' & (i\hm J_{\nu+1}(i\hm))' \\
 \mathcal{I}^+(\hy,\hx)    & J_\nu(\hx)    &  \hx   J_{\nu+1}(\hx)
\end{array}\right|
\left|\begin{array}{ccc}
 -\mathcal{I}^0(\hx,i\hm) & J_\nu(i\hm) & i\hm J_{\nu+1}(i\hm) \\
 -\mathcal{I}^0(\hx,i\hm)' & J_\nu(i\hm)' & (i\hm J_{\nu+1}(i\hm))' \\
 \tilde{\mathcal{I}}^-(\hx,\hy) & e^{-\hd^2/2}J_\nu(\hy) &
 e^{-\hd^2/2}\hy J_{\nu+1}(\hy)
\end{array}\right|
}{
\left|\begin{array}{cc}
 J_\nu(i\hm) & i\hm J_{\nu+1}(i\hm)\\
 J_{\nu-1}(i\hm) & i\hm J_{\nu}(i\hm)
\end{array}\right|^2
}~,
\label{eq:rho2connpqdeg}
\ee
The prime indicates the derivative with respect to $i\hm$. This formula is the limit of a 
more general
one with two non degenerate quark masses.
$\hx$ and $\hy$ are the rescaled eigenmodes for $\mu=0$ and
 $\mu=\delta$ respectively. 
The integrals are
\beea
{\cal I}^0(\hat{x},\hat{y}) &\equiv&
 \frac12 \int_0^1 dt\, J_{\nu}(\hat{x}\sqrt{t})J_{\nu}(\hat{y}\sqrt{t})
=
 \frac{\hat{x}J_{\nu+1}(\hat{x})J_{\nu}(\hat{y}) - \hat{y}J_{\nu+1}(\hat{y})
J_{\nu}(\hat{x})}{\hat{x}^2-\hat{y}^2} \cr
{\cal I}^\pm(\hat{x},\hat{y}) &\equiv& \frac12
 \int_0^1 dt \, e^{\pm\hd^2 t/2}J_{\nu}(\hat{x}\sqrt{t})J_{\nu}(\hat{y}\sqrt{t}) ~.
\label{eq:shortint}
\eea

Lattice data is never as beautiful as an analytic formula. In our case,
the problems we face in applying Eq.~\ref{eq:rho2connpqdeg} to 
simulation data are that
\begin{itemize}
\item
We have limited statistics.
\item
We do not compute a large number of eigenmodes.
\item
From our analysis of the integrated eigenvalue spectra
we suspect that cutoff effects modify the spectrum of the higher modes
 away from the RMT prediction.
\item
Simulations are done in finite volume.
\end{itemize}
Each of the four points contributes to the systematic or statistical error of our final result. 

\begin{figure}
\begin{center}
\includegraphics[width=0.5\textwidth,clip]{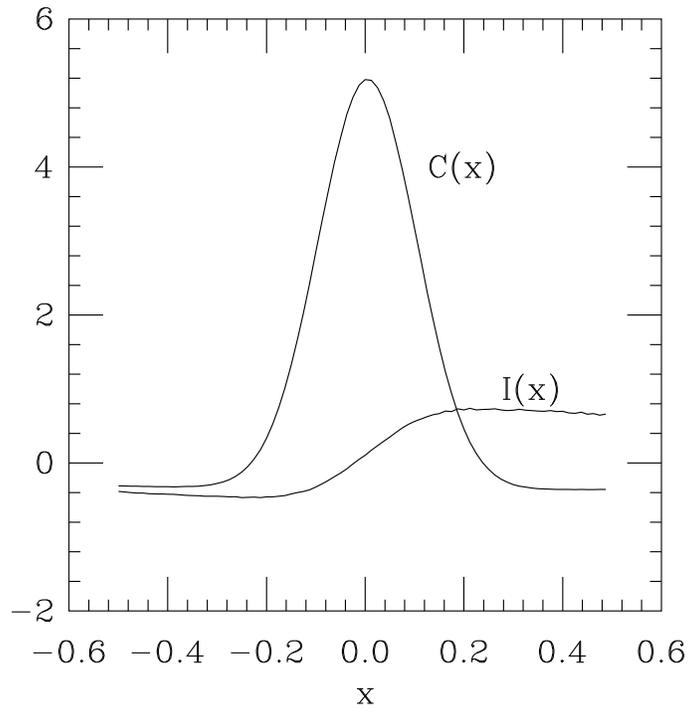}
\end{center}
\caption{$C(x)$ and $I(X)$ from Eqs. \protect{\ref{eq:cx}} and \protect{\ref{eq:ix}},
 for $\mu= m \Sigma V=2$, $\zeta_{max}\Sigma V=7$, $\nu=0$.
\label{fig:rmtint}
}
\end{figure}

The first problem on our list is the {\it limited statistics}.
It makes the comparison of the measured eigenvalue data to the theoretical
distribution of Eq.~\ref{eq:rho2connpqdeg} difficult. It is
therefore convenient to compute integrated correlators: The one suggested by
Refs. \cite{Damgaard:2005ys} and \cite{Damgaard:2006pu} is
\bee
C(x,\zeta_{max})=\int_0^{\zeta_{max}} dy \rho(x+y,y) \ .
\label{eq:cx}
\ee
To avoid binning the data, which is highly ambiguous given the low statistics, 
we perform a second integral and take instead
\bee
I(X,\zeta_{max})= \int_{-\zeta_{max}}^X C(x,\zeta_{max})dx \ .
\label{eq:ix}
\ee
The first integral $C(x)$ shows a spike at $x=0$ whose width goes
roughly as $\delta^2$. Then $I(X)$ will show a sharp step at $X=0$.
This is illustrated in Fig. \ref{fig:rmtint}.
(In practice, we have only studied $I(X)$.)

\begin{figure}
\begin{center}
\includegraphics[width=0.8\textwidth,clip]{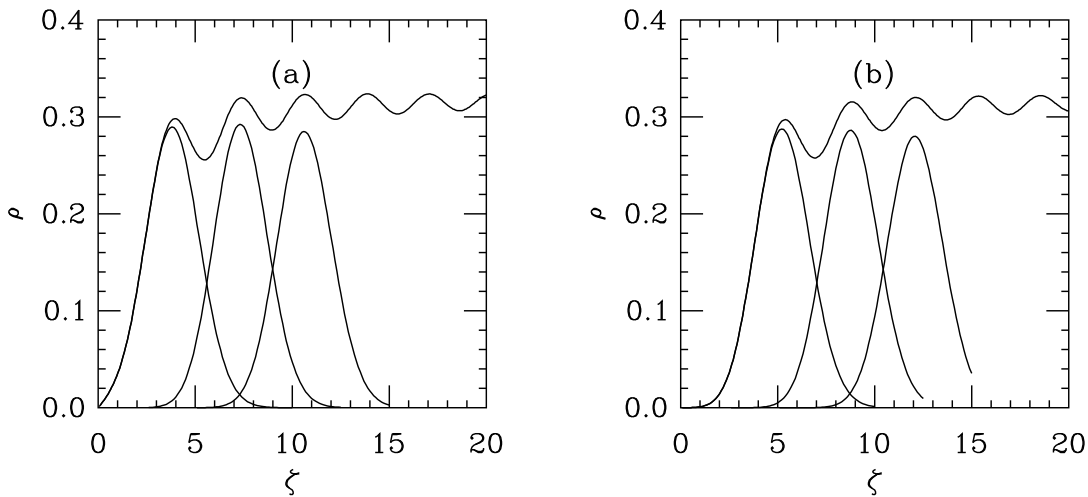}
\end{center}
\caption{An illustration of the RMT eigenvalue density (for all modes) and
the distributions of single eigenvalues, for $\mu= m \Sigma V=2$
as a function of $\hat\zeta=\lambda \Sigma V$, in topological sectors (a) $\nu=0$, (b) $|\nu|=1$.
\label{fig:rmtev}
}
\end{figure}

To deal with the second and third problems, we realize that for any finite
$\zeta_{max}$, only a {\it finite number  of eigenmodes} contribute to
$C(x)$ or $I(X)$. Since we will be measuring the eigenvalue correlations
at tiny $\delta$, a plot of the individual eigenvalue distributions
will easily reveal which modes saturate the correlator below a certain
value $\zeta_{\rm max}$ of $\lambda$.
An example is illustrated in Fig. \ref{fig:rmtev}. In this case
$m\Sigma V=2$. We see that for $\zeta_{max} \Sigma V \le 7$ only the two lowest modes
contribute to the $\nu=0$ correlator, and similarly for $\zeta_{max} \Sigma V =8$
 for $|\nu|=1$.

Let us now come to the fourth problem.
{\it Finite volume} typically modifies field theory correlation functions, because
the propagators are altered by the boundary conditions. In the epsilon
regime, finite volume also suppresses long-range order. To compute the correction to the
condensate, one basically writes the Goldstone boson fields as  $U(x) = U_0
\exp(i\pi(x)/F)$,
where $U_0$ is the zero mode field which enters into all infinite-volume
correlators (and is the variable connected to RMT formulas) and
$\pi(x)$ is small ($\sim 1/L^{d+2}$ in $d-$dimensions). Then one expands the partition
function to desired order in $1/L$ and integrates out the small terms.
For the condensate, the effect is to replace $\Sigma$ by $\Sigma_L = \rho_\Sigma \Sigma$
where
\bee
\rho_\Sigma = 1 + C_\Sigma \frac{1}{F^2} \Delta(0) +\dots
\ee
with $\Delta(0)$  the contribution to the tadpole graph (propagator
at zero separation) from finite-volume image terms. In the epsilon
regime, $\Delta(0) = -D/\sqrt{V}$
and $D$ depends on the geometry\cite{Gasser:1986vb}. (It is 0.1405 for
hypercubes.)  For $N_f$ flavors, $C_\Sigma = -(N_f^2-1)/N_f$.

We are aware of no equivalent calculation appropriate to the case in hand.
It amounts to computing corrections to derivatives of the partition function in the
 presence of a static
background isospin vector potential. It includes both physical quarks
 (or pseudoscalar mesons) with ghost states.

However, at the end of the day, we want to make a comparison with phenomenology.
We will make a guess (readers who do not like plausibility arguments are invited to look away for two paragraphs):

The epsilon-regime correction to $\Sigma$ is the same as the one-loop correction in the p-regime,
with the same $C_\Sigma$. All that differentiates the two is the evaluation of $\Delta(0)$.
We will just assume that the same result obtains for $F$ -- that
\bee
F_L = \rho_F F
\ee
where $\rho_F$ is the analog of $\rho_L$ with $C_F= - N_f/2$ replacing $C_\Sigma$.

Is this a plausible guess? Epsilon regime axial vector and vector current correlators
in boxes of length $T$
have been computed by several authors\cite{bunchofcites}.
The mass-independent part of the correlator is a constant, 
\bee
C(t) \sim \frac{F^2}{2T}\left[ 1 + \frac{N_f}{F^2}\left(\frac{D}{\sqrt{V}} - k\frac{T^2}{V} \right) \right].
\ee
In a symmetric box ($V=T^4$), $k=D/2$ and the effect is equivalent to $F_L = \rho_F F$
with $C_F= - N_F/(2\sqrt{2})$. Our guess scales properly with volume, but has a different
factor. There is an additional $1/\sqrt{V}$ correction in $C(t)$, which carries
a nontrivial $t$ dependence. It might be, that the effect of finite volume is to modify
the functional dependence of Eq. \ref{eq:rho2connpqdeg}.
Clearly, we have exceeded the bounds of the numerical simulation we set out to do.

\section{Numerical Simulation}
Let us now come to the discussion of the actual simulation.
Finding eigenmodes at nonzero isospin is easy: take a set of gauge configurations
from the dynamical stream, rotate the time-like links
\bee
U_0(x) \rightarrow \exp(i \mu) U_0(x)
\label{eq:rotate}
\ee
and re-compute eigenmodes of the massless Dirac operator.
This (standard \cite{Hasenfratz:1983ba}) implementation insures
 that the chemical potential -- and hence our extracted $F$ --
 is not renormalized by lattice interactions.

Our version of the overlap uses various implementations of fat links, but as Eq. \ref{eq:rotate}
 is a global rotation, the fat links
are rotated by the same amount. (That is, we first fatten the links,
then rotate them.) And these quenched measurements of eigenvalues
 are basically ``for free,'' as compared
to the cost of the generation of the configurations.

Our data set uses a lattice volume of $12^4$ points. 
The lattice spacing $a$, as determined from the Sommer parameter $r_0$ \cite{Sommer:1993ce},
is $r_0/a=3.71(5)$.  This calculation is ``parasitic'';
 our main research program with our simulations involves heavier quark masses.
 We present data from the lightest quark mass at which we have
extensive statistics, $am_q=0.03$. The pseudoscalar mass is
$am_\pi=0.329(3)$. This is not in the epsilon regime, but experience shows RMT is
robust
enough to work well outside the epsilon regime, and anyway, we are just making a
pilot study. In our conversions to physical
 units we use $r_0=0.5$ fm.

The overlap operator uses a ``kernel action'' (the nonchiral
 action inserted in the usual overlap formula) with nearest and
 next-nearest (diagonal) neighbors. The
$12^4$ data set uses the differentiable hypercubic smeared link of
 Ref. \cite{Hasenfratz:2007rf}. 
Details of the actions are described in
 Refs.
\cite{DeGrand:2006nv,DeGrand:2000tf,DeGrand:2004nq,DeGrand:2006ws};
everything except the choice of smearing for the link is identical to previous work. 

Let us briefly summarize the algorithmic set-up of our computation. 
For the $N_f=2$ simulations we use the reflection/refraction algorithm
first devised in Ref.~\cite{Fodor:2003bh}. In order to improve the tunneling
rate and precondition the fermion determinant we use two additional heavy 
pseudo-fermion fields as suggested by Hasenbusch\cite{Hasenbusch:2001ne}.
The integration is done with multiple-time scales\cite{Urbach:2005ji}. The runs 
for all sea quark masses were performed within a few months 
on a cluster of 32 Opteron CPU's which are connected by an Infiniband network.
For the $am_q=0.03$ ensemble, we collected about 400 thermalized HMC trajectories of unit length 
and analyzed 30 $\nu=0$ lattices and 75 $|\nu|=1$ ones.
We are computing eigenmodes using the ``Primme'' package of
McCombs and Stathopolous\cite{primme}.

\begin{figure}
\begin{center}
\includegraphics[width=0.4\textwidth,clip]{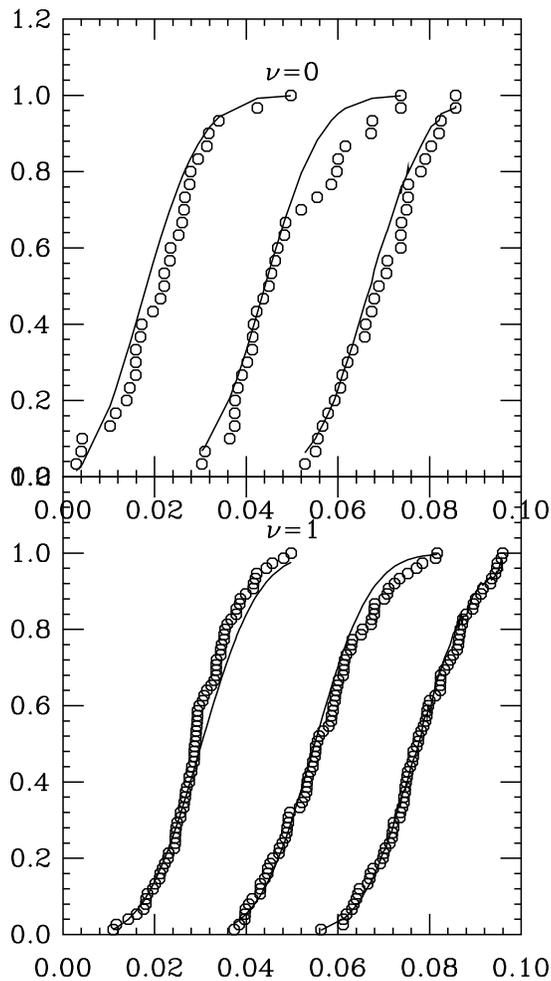}
\end{center}
\caption{Cumulant distribution for the three lowest 
eigenmodes (without chemical potential) in the $|nu|=0$ and 1 sectors, for our
 $12^4$ $\beta=7.3$ $am_q=0.03$ data set. The curve is a fit to
the lowest mode in $\nu=0$ and in $|\nu|=1$, with $a^3\Sigma_L=0.0073$. 
\label{fig:exdata}
}
\end{figure}

Our analysis begins  with a determination of $a^3 \Sigma_L$ by fitting
the RMT prediction of the distribution of the lowest eigenvalues of the 
Dirac operator without chemical potential to the data. 
As mentioned in the introduction, this method has been used by us and other groups during the last years 
and will serve as a cross check for the new method.
Specifically, we fit  the integrated eigenvalue density  (cumulant)
\bee
P_i(\lambda)=\int_0^\lambda p_i(\lambda') {\rm d}\lambda'
\ee
with $p_i(\lambda)$ the density of the $i$-th lowest eigenmode.
An example of a  typical fit, this one to the lowest eigenmode in the 
$\nu=0$ and $|\nu|=1$ sectors, is
 shown in 
Fig. \ref{fig:exdata}. Its best fit value is $a^3 \Sigma_L = 0.0073(3)$,
the quality factors for each individual mode from the combined fit are listed in Table~\ref{tab:qual}.
Fits to other sets of eigenmodes produce entirely consistent results.

\begin{table}
\begin{tabular}{c|c|ccc}
\ \ $|\nu|$\ \ &\ \ N \ \ &\ \  mode 1\ \  &\ \  mode 2\ \ &\ \  mode 3 \\
\hline
0    & 30  & 0.34 & 0.30 & 0.31\\
1    & 75  & 0.29 & 0.33 & 0.53 
\end{tabular}
\caption{The quality factor of the individual modes for the result of a combined
RMT fit to the lowest mode in each topological sector,
see Fig.~\ref{fig:exdata}.\label{tab:qual}}
\end{table}

Now we proceed to the analysis of the eigenvalues with chemical potential and fits
to Eq.~\ref{eq:ix}.
The correlator depends on two free parameters, $\hd$ and the scaled condensate
 (times volume) $\Sigma V$.
The ``experimental'' $I(X)$ varies almost continuously in $X$ when there are many
 lattices in the data set. Since it is an integral, its values at different $X$ are
highly correlated. 
We account for correlations via a bootstrap analysis:
We pick a bin size for $X$ and make a set of bootstrap-averaged
determinations of $I(X_j,\zeta_{max})$, with an uncertainty
 of a bin value $\sigma_j$  determined from the bootstrap.
The we do a fit minimizing  
\bee
\chi^2 = \frac{1}{N_b-2}\sum_{j=1}^{N_b} \frac{(I(x_j)- I(x_j,\hd,\Sigma V))^2}{\sigma_j^2}.
\label{eq:chisq}
\ee
We determine uncertainties in the fit parameters from the fluctuations in a bootstrap
average over a number of these fits.
The fits involve a number of arbitrary choices, and our results should be independent of them.
We have tested all of the following
\begin{itemize}
\item{Range of $\zeta_{max}$ }
\item{Range of $X$. }
\item{Number of values of $X_j$}
\end{itemize}

The fit then determines a favored value of $\hd$, from which $F_L= \hd/(\sqrt{V}\mu)$.
It also gives a second determination of the condensate $\Sigma_L V$,
to be compared to the previously discussed  direct fit to cumulant distributions.

\begin{figure}
\begin{center}
\includegraphics[width=0.4\textwidth,clip]{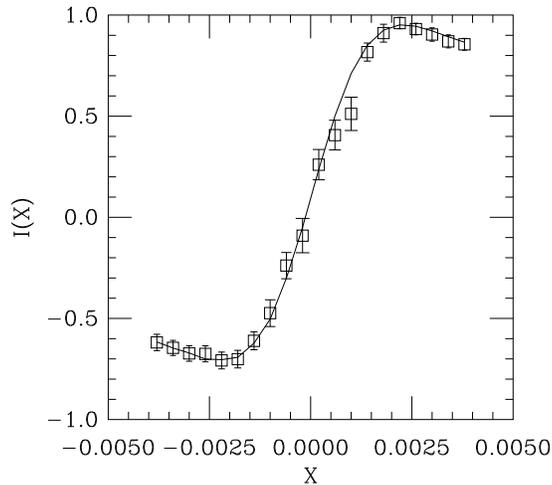}
\end{center}
\caption{Data (squares) and fit
 from
$|\nu|=1$. A cut $\zeta_{max}=0.07$ is enforced. The best fit values for this bootstrap
sample
are $\Sigma_L V=147$ and $aF_L=0.071$.
\label{fig:exstd}
}
\end{figure}

\begin{figure}
\begin{center}
\includegraphics[width=0.4\textwidth,clip]{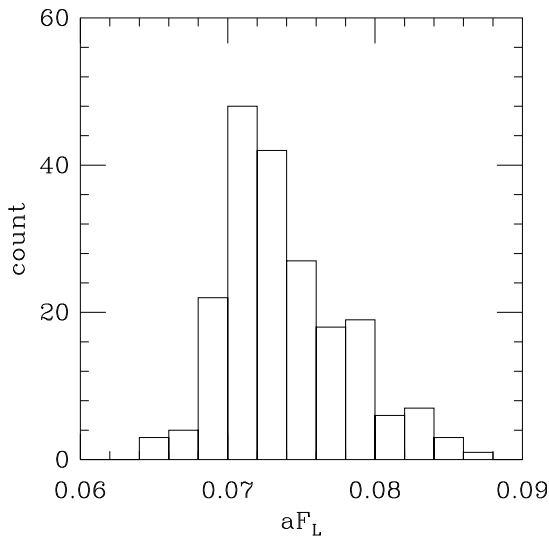}
\end{center}
\caption{Histogram of 200 bootstrap fits to $aF_L$ from the $|\nu|=1$ data set:
$\zeta_{max}=0.08$ and $X=0.004$ in the fits
\label{fig:hist}
}
\end{figure}

The cumulant distribution of the eigenmodes tells us
 how many modes contribute for any $\zeta_{max}$. In this case, we believe that we
can keep up to three eigenmodes and be consistent with random matrix theory
predictions, so a maximum value will be someplace near the average value of the
third eigenmode. This corresponds to a (not rescaled) $\zeta_{max}=0.07$ for
$\nu=0$ and 0.08 for $|\nu|=1$.

\begin{figure}
\begin{center}
\includegraphics[width=0.4\textwidth,clip]{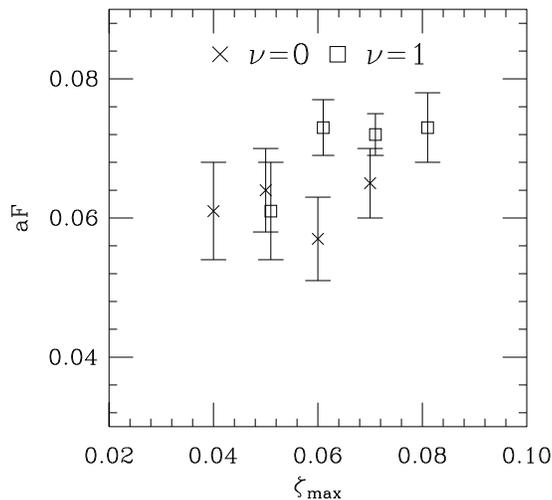}
\end{center}
\caption{Best-fit values of $aF_L$, varying $\zeta_{max}$.
\label{fig:comparezetas}
}
\end{figure}

Fig.~\ref{fig:comparezetas} shows the dependence of fits on $\zeta_{max}$.
These fits all include only the three lowest (but nonzero) eigenvalues in a topological sector.
Smaller  $\zeta_{max}$ means that less data is included in the fit; larger
 $\zeta_{max}$ means that more eigenmodes are needed to saturate the correlation function.
It appears that results
for this data set do not depend on the choice of  $\zeta_{max}$.

Finally, a histogram of bootstrap values of the finite-volume decay constant $F_L$
is shown in Fig. \ref{fig:hist}.

As the best-fit values do not show strong dependence on $\zeta_{max}$ we choose
to quote results from the $\zeta_{max}=0.08$ fit  for the $|\nu|=1$ data set and
$\zeta_{max}=0.07$ for $\nu=0$.
They give
$aF_L=0.065(5)$ and $a^3 \Sigma_L=0.0068(2)$; the $|\nu|=1$ set gives
$aF_L=0.073(5)$ and  $a^3 \Sigma_L=0.0072(2)$. Combining these fits
gives $aF_L=0.069(4)$ and $a^3 \Sigma_L=0.0071(1)$.
The fit gives a corroborating result for lattice-regulated $\Sigma$, consistent with
our result from eigenmode cumulants.

To complete the calculation of $\Sigma$, we need to convert from lattice regularization
to $\overline{MS}$. We do this using the intermediary of the RI-MOM scheme
\cite{Martinelli:1994ty}. Our methodology is basically identical to 
what we did in Ref.~\cite{DeGrand:2006nv}. 

\begin{figure}
\includegraphics[width=0.8\textwidth]{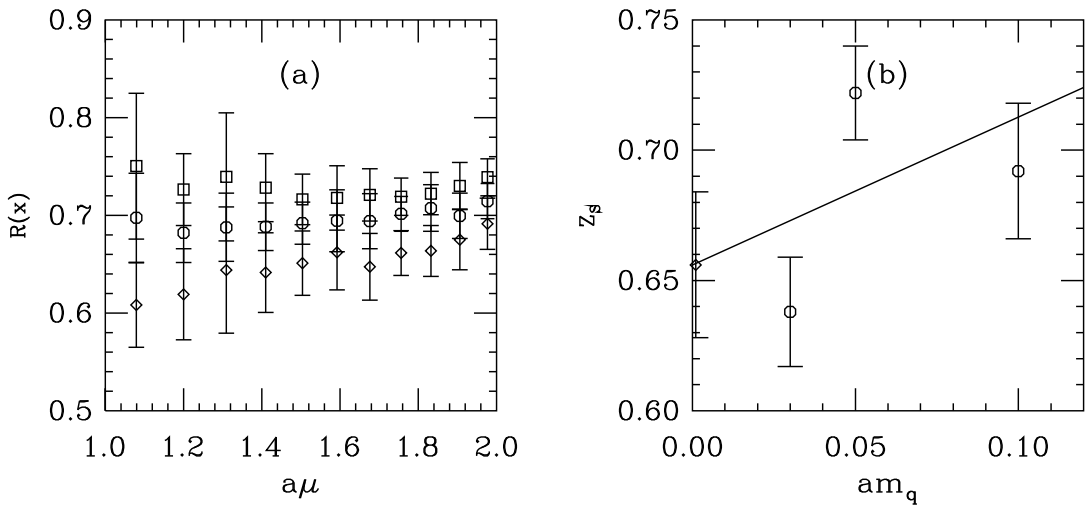}
\caption{\label{fig:extrap}
(a) $Z_S$ from our RI-MOM analysis:
octagons are $am_q=0.10$, squares, 0.05 and diamonds
0.03. Errors are from a jackknife average.
(b) $Z_S(RI')$ interpolated to $\mu=2$ GeV, vs $am_q$. The diamond shows the extrapolated point
(with its error).}
\end{figure}

We collected data from the three quark masses, $am_q$ = 0.10, 0.05, and 0.03.
We took  40, 30, and 40 lattices, respectively.
 In
RI-MOM, a  Z-factor is basically
the product of an averaged unamputated momentum-space
vertex function times two
averaged inverse propagators, so for an N-lattice data set we formed N
single-elimination jackknifed vertex-propagator combinations.
The jackknife-averaged Z-factors are shown in  Fig. \ref{fig:extrap}(a).

We want the Z-factor at a fiducial scale $\mu$ which conventionally is
chosen to be 2 GeV. We fit the Z-factor for a particular mass to a linear
dependence in $\mu$ for $\mu$'s ``nearby'' $\mu=2$ GeV.
There is one value for each possible value of lattice momentum.
Combining all the points with the same value of $\sum_i k_i^2$
gives a number with an error bar for a given value of $\sum_i k_i^2$
for each of these jackknifed data sets.

We interpolate these fits to the desired value of $\mu$,
 2 GeV with the lattice spacing taken from our Sommer parameter measurements. The
uncertainty comes from jackknifing these results. For our simulations,
 $\mu=2$ GeV corresponds to $a\mu \simeq 1.4$, and we did fits over the
range 1.0 or 1.1 to 1.6. (To be precise, 1.497 at $am_q=0.10$, 1.45 at 0.05,
1.368 at 0.03.)
The interpolated results are shown in Fig. \ref{fig:extrap}(b). A linear fit
gives the $am_q=0$ Z-factor of 0.656(28).

To convert to $\overline{MS}$ scheme, we
use the coupling constant from the so-called ``$\alpha_V$" scheme.
The one-loop
expression relating the plaquette to the coupling is
\begin{equation}
\ln\frac{1}{3}\Tr U_p=-\frac{8\pi}{3}\alpha_V(q^*)W,
\end{equation}
where $W=0.366$ and $q^*a=3.32$ for the tree-level L\"uscher--Weisz
action. All three data sets gave $\alpha_V(q^*)=0.188$. The lattice spacing
does not vary much in the three data sets, so converting any of them gives
$\alpha_s^{\overline{MS}}$(2 GeV)$=0.207$ and
$Z_S^{\overline{MS}}/Z_S^{RI'}=1.158$. 
Thus the combined lattice to $\overline{MS}$
conversion factor is $Z_S^{\overline{MS}}$(2~GeV)=0.76(3).
We insert our lattice determination for $aF_L$ in $\rho_\Sigma$. Dividing it out,
we find 
\bee
r_0 \Sigma (\overline{MS},\mu=2 \ {\rm GeV})^{1/3} = 0.594(13) 
\ee
or
\bee
( \Sigma(\overline{MS},\mu=2 \ {\rm GeV}) )^{1/3} = 234(4) {\rm MeV}
\ee
From the cumulant analysis, $r_0 \Sigma (\overline{MS},\mu=2 \ {\rm GeV})^{1/3} =0.600(13)$
or $( \Sigma(\overline{MS},\mu=2 \ {\rm GeV}) )^{1/3} = 237(6)$ MeV.
This seems to be quite consistent with other recent 
 determinations \cite{McNeile:2005pd,DeGrand:2006nv,Lang:2006ab,Fukaya:2007fb}.

Now for $F$. Combining results for $\nu=0$ and $|\nu|=1$, we find
\bee
r_0 F_L = 0.255(13) 
\ee
or with $r_0=0.5$ fm,
\bee
F_L = 101(6) {\rm MeV}.
\ee

This is a little high compared to the phenomenological
estimates of Refs. \cite{Gasser:1983yg,Colangelo:2003hf}.
 There are ample reasons to suspect that some of
this discrepancy could be due to fundamental problems with the simulation
($N_f=2$ is not real world), but we have already argued
that we expect that $F_L$ differs from $F$, by a factor which differs from unity by $O(1/(F^2L^2)$.

Using the guess for finite value corrections, $F_L = \rho_L F$, and our guess for
$\rho_L$, we estimate
\bee
r_0 F = 0.213(11) 
\ee
or
\bee
F = 84(5) {\rm MeV}.
\ee

Perhaps the error is more interesting than the central value:
We have produced a six per cent statistical uncertainty for a full QCD
quantity, with an extremely modest investment of computer power.

We remarked earlier that large scale simulations using asymptotic correlators typically produce
calculations of $f_\pi$ with small uncertainties but noisier predictions for $F$.
As an example, 
MILC\cite{Aubin:2004fs} quotes
(their Table IV, Fit A, our conversion) $r_1 F =0.131(10)$ or  $F=82(6)$ MeV.
($r_1$ is defined through the 
heavy-quark force as $r_1^2F(r_1)=1.0$.)
ETMC gets $r_0 F=0.197(2)$ \cite{Boucaud:2007uk}.
We believe that our result used several orders
of magnitude less computer power to produce a comparably-sized statistical error. Control 
over the systematic error due to the finite volume, however, remains a topic of future research.

\section{Conclusion}
This was a pilot study, and it needs several ingredients before a complete
calculation of $F$ can be performed:
\begin{itemize}
\item
A real analytic calculation of the finite volume correction to the  eigenvalue correlator 
is needed. Is it of the form $F_L= \rho_L F$ as we assumed, and if so, what is $\rho_F$?
In the absence of such a calculation, simulations at several volumes will be needed.
\item
A real numerical calculation needs to truly enter the epsilon regime, through a combination
of lower pseudoscalar mass and larger volume.
\item
The best way we know to extract the condensate from Dirac eigenmodes used
the distribution of individual eigenmodes, not the sum over many modes.
 The analog for $F$ would
be a calculation of the correlator of one eigenmode of ordinary fermions
 with one eigenmode
of partially quenched fermions in an isospin chemical potential.
We understand \cite{PDpc} that such a calculation is in progress.
\item
Finally, while we said at the beginning of this paper that $F$ apparently
 does not need
to be a supercomputing project, finding the physical decay constants
 involves determining
higher-order coefficients in the chiral Lagrangian. At present, this
 can only be done
in a spectroscopy-type calculation, from the asymptotic behavior of a
 correlation function.
This can be expensive.
Is there a way \cite{Akemann:2003tv} to remove
that bottleneck by relating these coefficients to some property of Dirac eigenmodes?
\end{itemize}

\section*{Acknowledgments}

This work was supported in part by the US Department of Energy.
T.~D. would like to thank Poul Damgaard for encouraging him to attempt 
this calculation
and for much valuable advice.
He would also like to thank Kim Splittorf for clarifying correspondence.
We would like to thank Hidenori Fukaya and
Silvia Necco for their comments on finite volume corrections.


\begin{thebibliography}{99}

\bibitem{Shuryak:1992pi}
E.~V.~Shuryak and J.~J.~M.~Verbaarschot,
Nucl.\ Phys.\ A {\bf 560}, 306 (1993) [arXiv:hep-th/9212088].

\bibitem{Verbaarschot:1993pm}
J.~J.~M.~Verbaarschot and I.~Zahed,
Phys.\ Rev.\ Lett.\  {\bf 70}, 3852 (1993) [arXiv:hep-th/9303012].

\bibitem{Verbaarschot:1994qf}
J.~J.~M.~Verbaarschot,
Phys.\ Rev.\ Lett.\  {\bf 72}, 2531 (1994) [arXiv:hep-th/9401059].


\bibitem{Berbenni-Bitsch:1997tx}
M.~E.~Berbenni-Bitsch, S.~Meyer, A.~Sch\"afer, J.~J.~M.~Verbaarschot and T.~Wettig,
Phys.\ Rev.\ Lett.\  {\bf 80}, 1146 (1998) [arXiv:hep-lat/9704018].



\bibitem{Damgaard:1998ie}
P.~H.~Damgaard, U.~M.~Heller and A.~Krasnitz,
Phys.\ Lett.\ B {\bf 445}, 366 (1999) [arXiv:hep-lat/9810060].

\bibitem{Gockeler:1998jj}
M.~G\"ockeler, H.~Hehl, P.~E.~L.~Rakow, A.~Sch\"afer and T.~Wettig,
Phys.\ Rev.\ D {\bf 59}, 094503 (1999)
[arXiv:hep-lat/9811018].


\bibitem{Edwards:1999ra}
R.~G.~Edwards, U.~M.~Heller, J.~E.~Kiskis and R.~Narayanan,
Phys.\ Rev.\ Lett.\  {\bf 82}, 4188 (1999)
[arXiv:hep-th/9902117].

\bibitem{Giusti:2003gf}
L.~Giusti, M.~L\"uscher, P.~Weisz and H.~Wittig,
JHEP {\bf 0311}, 023 (2003) [arXiv:hep-lat/0309189].


\bibitem{Follana:2005km}
  E.~Follana, A.~Hart, C.~T.~H.~Davies and Q.~Mason  [HPQCD Collaboration],
  Phys.\ Rev.\ D {\bf 72}, 054501 (2005)
  [arXiv:hep-lat/0507011].


\bibitem{Berbenni-Bitsch:1998sy}
M.~E.~Berbenni-Bitsch, S.~Meyer and T.~Wettig,
Phys.\ Rev.\ D {\bf 58}, 071502 (1998)
[arXiv:hep-lat/9804030].

\bibitem{Damgaard:2000qt}
P.~H.~Damgaard, U.~M.~Heller, R.~Niclasen and K.~Rummukainen,
Phys.\ Lett.\ B {\bf 495}, 263 (2000) [arXiv:hep-lat/0007041].

\bibitem{DeGrand:2006uy}
  T.~DeGrand, R.~Hoffmann, S.~Schaefer and Z.~Liu,
  Phys.\ Rev.\ D {\bf 74}, 054501 (2006)
  [arXiv:hep-th/0605147].



\bibitem{DeGrand:2006nv}
  T.~DeGrand, Z.~Liu and S.~Schaefer,
  Phys.\ Rev.\  D {\bf 74}, 094504 (2006)
  [Erratum-ibid.\  D {\bf 74}, 099904 (2006)]
  [arXiv:hep-lat/0608019].


\bibitem{Lang:2006ab}
  C.~B.~Lang, P.~Majumdar and W.~Ortner,
  Phys.\ Lett.\  B {\bf 649}, 225 (2007)
  [arXiv:hep-lat/0611010].

\bibitem{Fukaya:2007fb}
  H.~Fukaya {\it et al.}  [JLQCD Collaboration],
  Phys.\ Rev.\ Lett.\  {\bf 98}, 172001 (2007)
  [arXiv:hep-lat/0702003].

\bibitem{Fukaya:2007yv}
  H.~Fukaya {\it et al.},
  arXiv:0705.3322 [hep-lat].

\bibitem{McNeile:2005pd}
	Compare the summaries in
  C.~McNeile,
  Phys.\ Lett.\ B {\bf 619}, 124 (2005)
  [arXiv:hep-lat/0504006].




\bibitem{Akemann:2006ru}
  G.~Akemann, P.~H.~Damgaard, J.~C.~Osborn and K.~Splittorff,
  Nucl.\ Phys.\  B {\bf 766}, 34 (2007)
  [arXiv:hep-th/0609059].

\bibitem{Gasser:1987ah}
  J.~Gasser and H.~Leutwyler,
  Phys.\ Lett.\  B {\bf 188}, 477 (1987).

\bibitem{Damgaard:2005ys}
  P.~H.~Damgaard, U.~M.~Heller, K.~Splittorff and B.~Svetitsky,
  Phys.\ Rev.\  D {\bf 72}, 091501 (2005)
  [arXiv:hep-lat/0508029].






\bibitem{Damgaard:2006pu}
  P.~H.~Damgaard, U.~M.~Heller, K.~Splittorff, B.~Svetitsky and D.~Toublan,
  Phys.\ Rev.\  D {\bf 73}, 074023 (2006)
  [arXiv:hep-lat/0602030].



\bibitem{Damgaard:2006rh}
  P.~H.~Damgaard, U.~M.~Heller, K.~Splittorff, B.~Svetitsky and D.~Toublan,
  Phys.\ Rev.\  D {\bf 73}, 105016 (2006)
  [arXiv:hep-th/0604054].



\bibitem{Neuberger:1997fp}
H.~Neuberger,
Phys.\ Lett.\ B {\bf 417}, 141 (1998)
[arXiv:hep-lat/9707022].

\bibitem{Neuberger:1998my}
H.~Neuberger,
Phys.\ Rev.\ Lett.\  {\bf 81}, 4060 (1998)
[arXiv:hep-lat/9806025].




\bibitem{Ginsparg:1981bj}
  P.~H.~Ginsparg and K.~G.~Wilson,
  Phys.\ Rev.\ D {\bf 25}, 2649 (1982).



\bibitem{Niedermayer:1998bi}
  F.~Niedermayer,
  Nucl.\ Phys.\ Proc.\ Suppl.\  {\bf 73}, 105 (1999)
  [arXiv:hep-lat/9810026].



\bibitem{Bernard:2006wx}
  C.~Bernard {\it et al.}  [MILC Collaboration],
  arXiv:hep-lat/0609053.



\bibitem{Gasser:1983yg}
  J.~Gasser and H.~Leutwyler,
  Annals Phys.\  {\bf 158}, 142 (1984).


\bibitem{Colangelo:2003hf}
  G.~Colangelo and S.~Durr,
  Eur.\ Phys.\ J.\  C {\bf 33}, 543 (2004)
  [arXiv:hep-lat/0311023].



\bibitem{Gasser:1986vb}
  J.~Gasser and H.~Leutwyler,
  Phys.\ Lett.\ B {\bf 184}, 83 (1987);
  Nucl.\ Phys.\ B {\bf 307}, 763 (1988).
See also
  P.~Hasenfratz and H.~Leutwyler,
  Nucl.\ Phys.\ B {\bf 343}, 241 (1990).


\bibitem{bunchofcites}
A partial citation path includes
  F.~C.~Hansen,
  Nucl.\ Phys.\  B {\bf 345}, 685 (1990);
  F.~C.~Hansen and H.~Leutwyler,
  Nucl.\ Phys.\  B {\bf 350}, 201 (1991);
  P.~H.~Damgaard, P.~Hernandez, K.~Jansen, M.~Laine and L.~Lellouch,
  Nucl.\ Phys.\  B {\bf 656}, 226 (2003)
  [arXiv:hep-lat/0211020];
  P.~Hernandez and M.~Laine,
  JHEP {\bf 0301}, 063 (2003)
  [arXiv:hep-lat/0212014].


\bibitem{Hasenfratz:1983ba}
  P.~Hasenfratz and F.~Karsch,
  Phys.\ Lett.\  B {\bf 125}, 308 (1983).


\bibitem{Sommer:1993ce}
  R.~Sommer,
  Nucl.\ Phys.\ B {\bf 411}, 839 (1994)
  [arXiv:hep-lat/9310022].

\bibitem{Hasenfratz:2007rf}
  A.~Hasenfratz, R.~Hoffmann and S.~Schaefer,
  JHEP {\bf 0705}, 029 (2007)
  [arXiv:hep-lat/0702028].



\bibitem{DeGrand:2000tf}
  T.~A.~DeGrand  [MILC collaboration],
  Phys.\ Rev.\  D {\bf 63}, 034503 (2001)
  [arXiv:hep-lat/0007046].


\bibitem{DeGrand:2004nq}
  T.~A.~DeGrand and S.~Schaefer,
  Phys.\ Rev.\  D {\bf 71}, 034507 (2005)
  [arXiv:hep-lat/0412005].


\bibitem{DeGrand:2006ws}
  T.~DeGrand and S.~Schaefer,
  JHEP {\bf 0607}, 020 (2006)
  [arXiv:hep-lat/0604015].


\bibitem{Fodor:2003bh}
  Z.~Fodor, S.~D.~Katz and K.~K.~Szabo,
  JHEP {\bf 0408}, 003 (2004)
  [arXiv:hep-lat/0311010].


\bibitem{Hasenbusch:2001ne}
  M.~Hasenbusch,
  Phys.\ Lett.\  B {\bf 519}, 177 (2001)
  [arXiv:hep-lat/0107019].

\bibitem{Urbach:2005ji}
 C.~Urbach, K.~Jansen, A.~Shindler and U.~Wenger,
 Comput.\ Phys.\ Commun.\  {\bf 174}, 87 (2006)
 [arXiv:hep-lat/0506011].

\bibitem{primme}
A. Stathopoulos, Nearly optimal preconditioned methods for hermitian
     eigenproblems under limited memory. Part I: Seeking one eigenvalue,
     Tech Report: WM-CS-2005-03, July, 2005. To appear in SIAM J. Sci. Comput.
A. Stathopoulos and J. R. McCombs, Nearly optimal preconditioned methods
     for hermitian eigenproblems under limited memory. Part II: Seeking many
     eigenvalues, Tech Report: WM-CS-2006-02, June, 2006.



\bibitem{Bernard:2000gd}
  C.~W.~Bernard {\it et al.},
  Phys.\ Rev.\  D {\bf 62}, 034503 (2000)
  [arXiv:hep-lat/0002028].






\bibitem{Martinelli:1994ty}
  G.~Martinelli, C.~Pittori, C.~T.~Sachrajda, M.~Testa and A.~Vladikas,
  Nucl.\ Phys.\ B {\bf 445}, 81 (1995)
  [arXiv:hep-lat/9411010].

\bibitem{Aubin:2004fs}
  C.~Aubin {\it et al.}  [MILC Collaboration],
  Phys.\ Rev.\  D {\bf 70}, 114501 (2004)
  [arXiv:hep-lat/0407028].

\bibitem{Boucaud:2007uk}
  Ph.~Boucaud {\it et al.}  [ETM Collaboration],
  Phys.\ Lett.\  B {\bf 650}, 304 (2007)
  [arXiv:hep-lat/0701012].



\bibitem{PDpc}
  G.~Akemann and P.~H.~Damgaard, private communication about work in progress.

\bibitem{Akemann:2003tv}
For an example of a related discussion, see
  G.~Akemann and P.~H.~Damgaard,
  Phys.\ Lett.\  B {\bf 583}, 199 (2004)
  [arXiv:hep-th/0311171].








\end{thebibliography}
\end{document}